\documentclass[aps,reprint,superscriptaddress]{revtex4-1}

\usepackage{graphicx}
\usepackage{dcolumn}
\usepackage{bm}
\usepackage{mathrsfs}
\usepackage[colorlinks]{hyperref}
\usepackage[T1]{fontenc}
\usepackage[utf8]{inputenx}
\usepackage{amsmath}
\usepackage{verbatim}
\usepackage{graphicx}
\usepackage{xcolor}

\usepackage{amsmath,comment}

\newtheorem{theorem}{Theorem}[section]

\def\d{{\, \rm d}}

\usepackage[nodisplayskipstretch]{setspace}

\begin{document}


\title{A Non-Gaussian Stochastic Model for the El Ni\~no Southern Oscillation }

\author{L. T. Giorgini}
\email{ludovico.giorgini@su.se}
\affiliation{Nordita, Royal Institute of Technology and Stockholm University, Stockholm 106 91, Sweden}

\author{W. Moon}
\email{woosok.moon@su.se}
\affiliation{Department of Mathematics, Stockholm University 106 91 Stockholm, Sweden}
\affiliation{Nordita, Royal Institute of Technology and Stockholm University, Stockholm 106 91, Sweden}

\author{N. Chen}
\email{chennan@math.wisc.edu}
\affiliation{Department of Mathematics, University of Wisconsin-Madison, Madison, Wisconsin 53706, USA}

\author{J. S. Wettlaufer}
\email{john.wettlaufer@yale.edu}
\affiliation{Yale University, New Haven, Connecticut 06520, USA}
\affiliation{Nordita, Royal Institute of Technology and Stockholm University, Stockholm 106 91, Sweden}

\date{\today}

\begin{abstract}
A non-autonomous stochastic dynamical model approach is developed to describe the seasonal to interannual variability of the El Ni\~no-Southern Oscillation (ENSO). We determine the model coefficients by systematic statistical estimations using partial observations involving only sea surface temperature data.  Our approach reproduces the observed seasonal phase locking and its uncertainty, as well as the highly non-Gaussian statistics of ENSO.  Finally, we recover the intermittent time series of the hidden processes, including the thermocline depth and the wind bursts.
\end{abstract}

\maketitle


Air-sea interactions in the tropical Pacific drive the largest interannual variability process in climate called the El Ni\~no-Southern Oscillation (ENSO), with an influence that reaches the higher latitudes via atmospheric and oceanic teleconnections \cite{McPhaden2006, GhilRMP}.  Due to its spatiotemporal impacts, estimating the seasonal to interannual state of ENSO is essential for predicting a wide range of regional and global climate phenomena \cite{Battisti1995, Havlin2015}.  However, the complexity of ENSO, involving as it does stochastic forcing from atmospheric transients \cite[e.g.,][]{Kleeman1997} and nonlinear air-sea interactions \cite[e.g.,][]{An2004}, poses great challenges in both modeling and prediction.  


There are two traditional methods of modeling ENSO \cite[see e.g.,][for a review]{Tang2018}. One is to use the state-of-art coupled general circulation models (GCMs), which solve the governing conservation (energy, mass, momentum) equations for the rotating planet with many sub-grid-scale parameterizations of unresolved small-scale processes.  The other is to build low-order statistical models.  The former treat the main physical aspects of ENSO and thus provide reasonable representations of the spatiotemporal patterns and air-sea interactions from intraseasonal to interannual time scales \cite{Jan2005}.  The latter, commonly designed to describe and predict particular ENSO indices, characterize the large-scale features of the system and are much cheaper computationally than GCMs \cite{Petrova2020}.  
Nonetheless, despite great progress in developing these methods, there remain limitations to advancing analysis and prediction.  On the one hand, GCMs incur systematic errors originating from inaccurate state estimations and incomplete sub-grid-scale parameterizations associated with the ocean thermocline, tropical instability waves, and the Madden-Julian Oscillation (MJO) \cite{Chen2008, Barnston2015}.  On the other hand, the statistical models based on multivariate regressions cannot provide detailed physical information regarding the phase and intensity of ENSO \cite{Tippet2019}.  One direction for progress is to utilize a hybrid strategy that simultaneously maximizes the advantages of both physically-oriented and statistical models.

The basic theoretical understanding of the tropical air-sea interactions underlying ENSO \cite{Cane1985, Jin1993} has lead to a hierarchy of simple models of the main processes that control the sea surface temperature (SST) in the tropical ocean \cite[e.g.,][]{Wang2017}.  A prominent approach is the recharge-discharge model \cite{Jin1}, and its extensions and generalizations, derived from the forced shallow water equations using the two-strip approximation.  Despite it being a two-dimensional linear model, it (and its generalizations) can capture inter-annual variability, seasonal variability with time-periodic coefficients \cite{Stein2010, Levine2015}, and a range of stochastic forcing representing weather and intraseasonal processes, such as westerly wind bursts (WWBs) \cite{Chen2017}.


Here we describe a two-stage stochastic model approach that captures the large-scale dynamical and statistical characteristics of ENSO.  
Stochastic recharge-discharge oscillator models have used observations to recover statistics on inter-annual time-scales \cite{Burgers1999, Burgers2005}, and have incorporated the seasonal cycle of the Bjerknes feedback to treat ENSO phase-locking and the spring predictability barrier \cite{Stein2010, Stein2011, Levine2015}.  
However, while our approach builds upon the recharge-discharge model, we incorporate a slowly-varying low-order deterministic component that systematically treats time-varying coefficients and multiplicative noise to accurately describe 
the central intraseasonal to interannual features of ENSO. 
In particular, our framework reproduces ENSO's observed seasonal phase locking, uncertainty and highly non-Gaussian statistics.
However, we clearly depart from multivariate regression statistical models that require observations of all state variables.  
Rather, our non-Gaussian model incorporates a subset of observations from which we can infer unobserved quantities and quantify their uncertainties using an efficient and exact data assimilation scheme, and can thereby 
accurately recover the difficult to observe thermocline depth and wind bursts.  Finally, by systematically determining time-dependent model coefficients, we precisely simulate the seasonal to inter-annual ENSO statistics. 

We begin with the following coupled model, 
\begin{equation}
\begin{split}
&\dot{x}(t) = a(t)x(t) + \omega(t)h(t) + N(t)\xi_x(t) ~~ \textrm{and} \\&
\dot{h}(t) = -\omega(t)x(t) + \lambda h(t) + \sigma \xi_h(t),
\label{model}
\end{split}\end{equation}
in which $x(t)$ is the averaged SST anomaly in the equatorial central-eastern Pacific (the so-called Ni\~no 3.4 region), and $h(t)$ is the thermocline depth, which is a surrogate of the heat content, averaged over the western tropical Pacific. The time-dependent function, $a(t)$, represents the seasonal evolution of the Bjerknes feedback and the noise amplitude, $N(t)$, captures the role of the relatively short time-scale forcing of the SST anomaly $x(t)$. Both $\xi_x(t)$ and $\xi_h(t)$ are independent Gaussian white noise processes. Finally, the other time-dependent function, $\omega(t)$, controls the coupling of tropical atmosphere and the upper ocean, which is one of the key parameters that determines the
quasi-oscillatory behavior of the dominant ENSO modes, and in traditional low-order models is treated as a constant \cite{Stein2010}.  Here, however, treating it as time-dependent facilitates a more accurate description of 
the intraseasonal and interannual statistics.


We emphasize the role of the seasonality of background fields in the tropical Pacific by making all the time dependent coefficients in \eqref{model} periodic on the annual cycle. We determine the functions $a(t)$ and $N(t)$ using the monthly statistics of the Ni\~no 3.4 index \cite{Moon2017}. The coefficient $a(t)$ is related to the relationship between the monthly variance $\langle x^2(t) \rangle$ and the covariance between two
adjacent months $\langle x(t)x(t+\Delta t)\rangle$. We use the derived $a(t)$ to estimate $N(t)$ from the statistics of the derived data; $y=x(t+\Delta t)-x(t)-a(t)x(t)\Delta t$.  It has been shown previously that $a(t)$ and $N(t)$ 
can be accurately estimated independent of the other coefficients (see supplementary material in \cite{Moon2017}).
The coupling function, $\omega(t)$, and the constants $\lambda$ and $\sigma$, are determined by an expectation-maximization algorithm \cite{Dempster1977,Sundberg1974,Sundberg1976}, which is a useful and efficient statistical estimation method involving incomplete data.  We note 
that only the Ni\~no 3.4 index data is used as the partial observation for model calibration, whereas the thermocline depth, $h(t)$, is assumed to be unknown since it is not directly available from satellite observations. Therefore, as detailed in Appendix \ref{App_B}, \ref{App_C}, we infer $h(t)$ together with unknown coefficients $\omega(t)$, $\lambda$, and $\sigma$ \cite{chen2020learning}. 


\begin{figure*}
\centering
\includegraphics[width=1.00\linewidth]{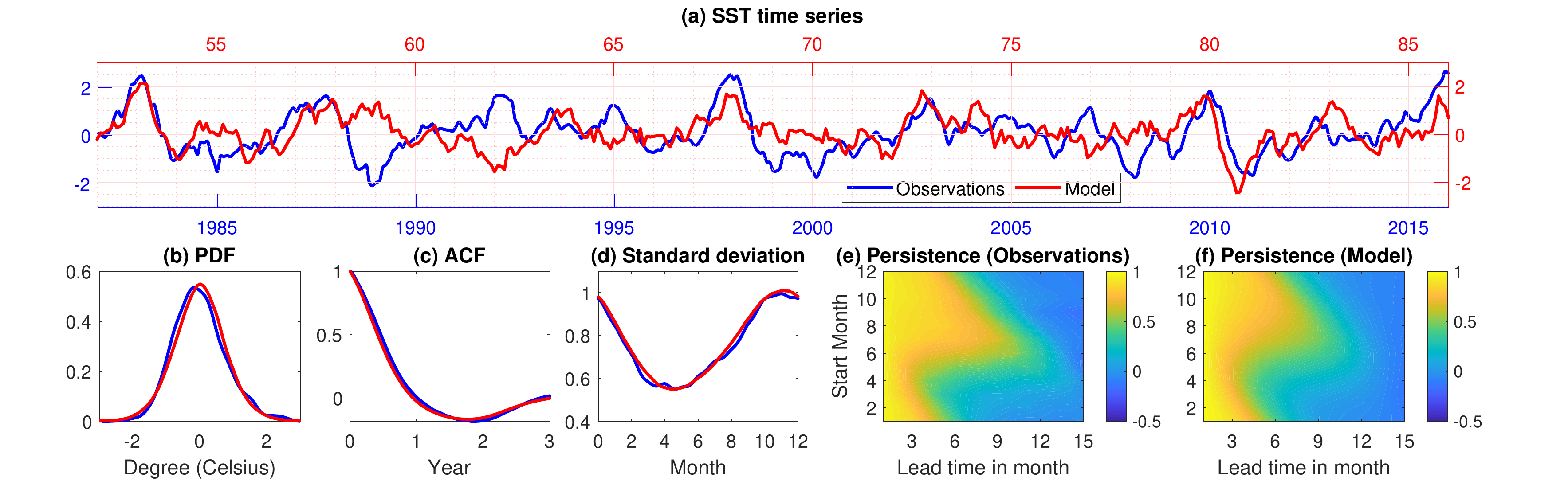}
\caption{Comparison of the observed SST (blue; bottom x-axis) with the model Eq. \eqref{model} (red; top x-axis). The SST time series are shown in (a), where the model result is one of the random realizations from  Eq. \eqref{model} that shares the same length as the observational data from 1983 to 2016. The comparisons of the PDFs and ACFs are shown in (b) and (c). The SST standard deviation as a function of calendar month is shown in (d). In (e) and (f), the diagram of the persistence forecast from the observational data and that from the model can be compared.
}
\label{Fig1}
\end{figure*}

We evaluate the quality of the model by  comparing the model statistics with those of the observed Ni\~no 3.4 index in Fig. \ref{Fig1}.  A qualitative comparison between a random realization 
of the model and the observed data show comparable variability in Fig. \ref{Fig1}(a) and quantitative comparisons of the probability density functions (PDFs)
and the autocorrelation functions (ACFs) of the SST anomaly shown in Figs. \ref{Fig1}(b) and (c) demonstrate that the model 
is able to accurately simulate the interannual quasi-oscillatory behavior of ENSO. 
Moreover, as seen in Fig. \ref{Fig1}(d), the model reproduces monthly standard deviations of the Ni\~no 3.4 index data with high precision.  Therefore, two important seasonal features of ENSO, ``phase locking'' to the seasonal cycle \cite{Tziperman1998} and the ``spring predictability barrier'' \cite{Levine2015}, are accurately captured in this two dimensional model. Phase locking is related to the activation of the positive Bjerknes feedback from summer to early winter, which leads to the concentration of abnormal ENSO events during November and December, as is clear from the maximum of the standard deviation near the end of the year. This phase locking feature is clearly seen from the model trajectory (Fig. \ref{Fig1}a), where the peaks of nearly all of the major ENSO events occur in the boreal winter, consistent with the observations. The spring predictability barrier is related to the increase of the noise magnitude, $N(t)$, during the spring, which increases model uncertainties during the boreal summer thereby causing the loss of predictability \cite{Moon2017}. Indeed, this is reflected by the structural similarity between the model and the observations in the persistence forecast diagrams shown in Figs. \ref{Fig1}(e) and (f).

Despite its simplicity, the model system of Eqs. \eqref{model} recovers the monthly standard deviation and ACFs of the Ni\~no 3.4 index with higher accuracy
than coupled GCMs \cite{Bellenger2014}, linear and nonlinear statistical models \cite{Kondrashov2005}, and simpler recharge-discharge based approaches \cite{Stein2010}. 
Our systematic statistical estimation of coefficients enable this stochastic model to regenerate the statistics of the Ni\~no 3.4 index. 
In particular, as shown in Fig. \ref{Fig1}(c), the precise representation of the ACFs beyond 2 years suggests that the overall dynamical features of the thermocline depth, $h(t)$, are well captured in this model. Variations in $h(t)$, and thus of the equatorial warm water volume (WWV) in the western tropical Pacific, influence the SST through vertical advection of temperature anomalies by mean upwelling, a process known as the ``thermocline feedback''. Thus, the thermocline depth acts as a precursor for the inter-annual prediction of SST anomalies \cite{Meinen2000}.


Although the linear model described above successfully captures the seasonal variability and the interannual large-scale dynamical features of ENSO, it only allows Gaussian PDFs, whereas the observed statistics are highly non-Gaussian.  Importantly, a model must include atmospheric wind burst forcing, which plays a critical role in generating the extreme ENSO events responsible for non-Gaussianity \cite{Thual2016,Chen2017}.   Therefore, we include wind bursts in the stochastic model of Eqs. \eqref{model} as follows, 
\begin{equation}\begin{split}
&\dot{x}(t) = a(t)x(t) + \omega(t)h(t) +\alpha_1 \tau(t)+ N(t)\xi_x(t), \\&
\dot{h}(t) = -\omega(t)x(t) + \lambda h(t) + \alpha_2 \tau(t)+\sigma \xi_h(t) ~~ \textrm{and} \\&
\dot{\tau}(t)=d_\tau \tau(t)+\rho(x) \xi_\tau(t),
\label{model_WB}
\end{split}\end{equation}
where we adopt a simple linear process with multiplicative noise to generate the stochastic wind bursts $\tau(t)$ that drive the SST and thermocline dynamics. The parameters in the wind burst equation are chosen to accurately reproduce the observed wind stress (Fig. \ref{Fig2} (d)-(f)), while the coefficients of Eqs. \eqref{model} were slightly modified to take into account the additional process in the model.
Despite the non-Gaussianity of the model, the thermocline depth and the wind bursts remain as Gaussian processes conditioned on the observed SST.  
Moreover, by using the partial observations that involve only the Ni\~no SST data, we have closed analytic formulae for solving the conditional distribution
of the unobserved thermocline depth and the wind bursts \cite[see][and Appendix \ref{App_B}]{chen2018conditional},
\begin{equation}\label{filter_soln}
p(h(t),\tau(t)|x(s),s\leq t)\sim \mathcal{N}(\mu_f(t),R_f(t)),
\end{equation}
where $\mu_f(t)$ and $R_f(t)$ are the so-called filter mean and filter covariance in data assimilation.

The filter estimate \eqref{filter_soln} of the non-Gaussian model \eqref{model_WB} can be used to carry out an online time series reconstruction of the unobserved variables and analyze the associated uncertainty, which can also serve as the forecast initialization. Note that the observational data for the equatorial heat content (the integrated WWV above the 20${}^o$C isotherm between 5N-5S, 120E to 80W \cite{Meinen2000}, which is equivalent to the thermocline depth $h(t)$ in \eqref{model_WB}) is only available since 1983 \footnote{The WWV values are derived from the temperature analyses of the Bureau National Operations Centre (BNOC) at the Australian Bureau of Meteorology \cite{Smith1995}}, whereas SST data is available from the pre-industrial era. Therefore, the available SST data can be incorporated into the model \eqref{model_WB} to reconstruct the thermocline data over a much longer period.


We first validate the accuracy of the recovered $h(t)$ and $\tau(t)$ in the period after 1983. Fig. \ref{Fig2} (a) shows the filter mean time series of the thermocline depth, $h(t)$, (red) from the non-Gaussian model \eqref{model_WB}, which compares favorably to the observational value (blue) with a pattern correlation that is nearly $0.9$. Moreover, the small error between the observed and the recovered time series lies approximately within one standard deviation of the filter estimated uncertainty (pink shaded region). In particular, as expected from the physical role of the WWBs, the extreme events are accurately recovered.  Similar conclusions are reached for the wind burst time series (Fig. \ref{Fig2}e). In addition to a skillful state estimation, the reconstructed wind burst time series has the further influence of filtering out the independent white noise in observations.
As shown in Figs. \ref{Fig2}(b, f), the model PDFs capture those of the observations for both the thermocline depth and the wind bursts, particularly the significant non-Gaussian skewness and kurtosis.  Moreover, similar accuracy is found in ACFs and the power spectra as shown Figs. \ref{Fig2}(c, d, g, h).  The success in recovering unobserved processes is due to the confluence of the model containing the essential physics and properly constrained stochastic forcing and coefficients.  


\begin{figure*}
\centering
\includegraphics[width=1.00\linewidth]{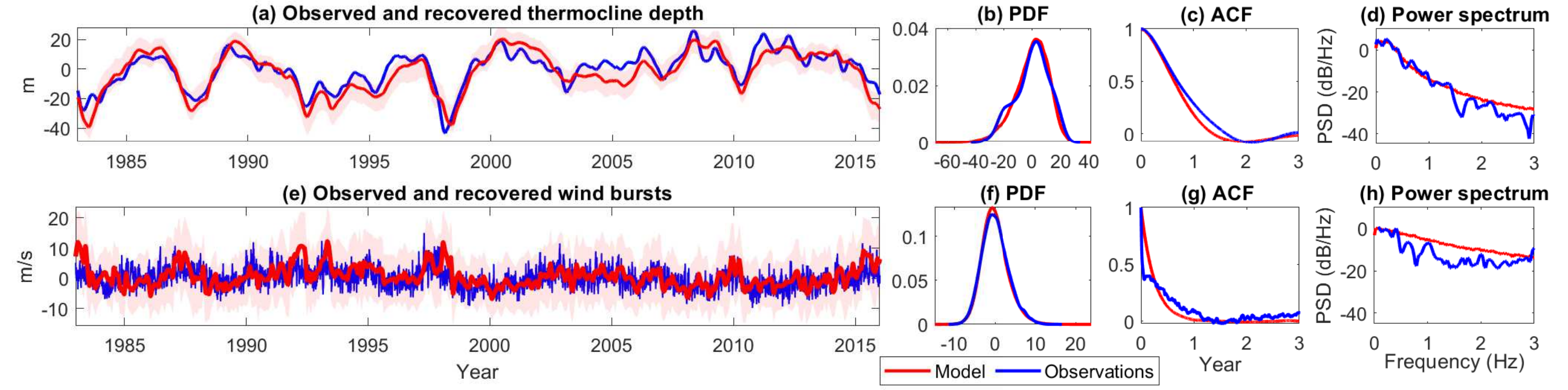}
\caption{Comparison of the observational value (blue) and the recovered time series from the filter estimate (red) using Eq. \eqref{filter_soln}, based on the non-Gaussian model \eqref{model_WB} for the thermocline depth (a) and the wind bursts (e). The solid red curves are the filter mean while the shaded red areas show one standard deviation in the filter estimate (the square root of the filter variance). The comparison of the PDFs, the ACFs and the power spectra between the observations and simulations from the non-Gaussian model \eqref{model_WB} are shown in (b)--(d), for the thermocline depth and (f)--(h), for the wind bursts.
}
\label{Fig2}
\end{figure*}

\begin{figure*}
\centering
\includegraphics[width=1.00 \linewidth]{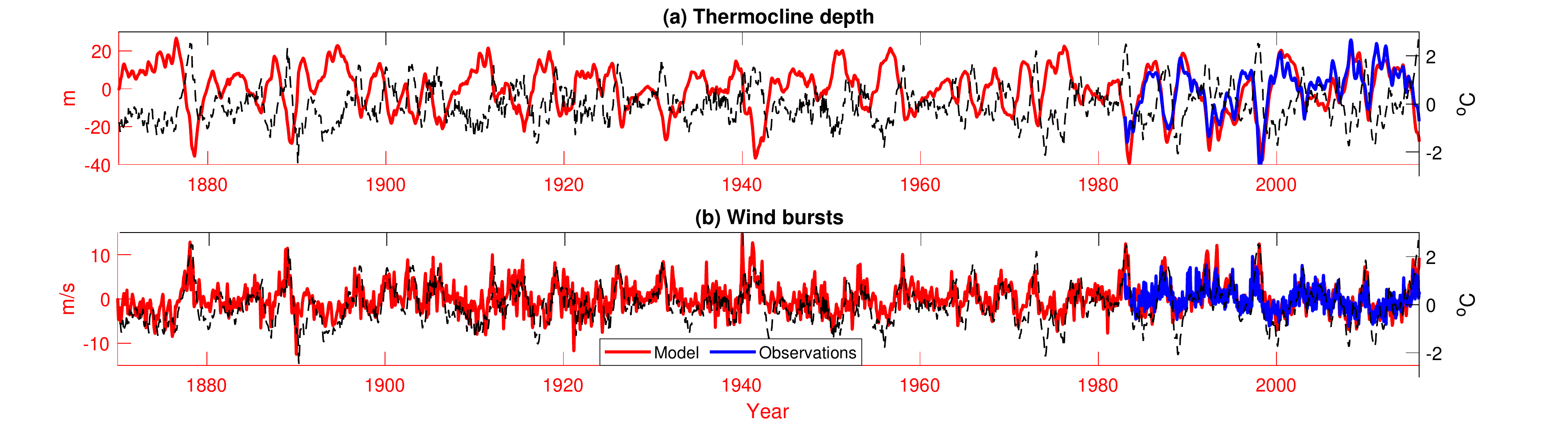}
\caption{The recovered time series of the hidden variables (red curves; left y-axis) for (a) the thermocline depth and (b) the wind bursts, using the non-Gaussian model, Eqs. \eqref{model_WB}, with the partial observations involving only SST data (dashed black curves; right y-axis) from 1870 to 2016. The method used here is the nonlinear smoother estimate, Eq. \eqref{smoother_soln}. The blue curves show the observational data from 1983 to 2016.}
\label{Fig3}
\end{figure*}

As shown Fig. \ref{Fig2}, the accurate nonlinear filter estimates of Eq. \eqref{filter_soln}  provide an online estimation method of the thermocline depth $h(t)$ and the wind bursts $\tau(t)$ relevant for model initialization. On the other hand, reconstructing historical data  can be achieved by solving a slightly different conditional distribution
\begin{equation}\label{smoother_soln}
  p(h(t),\tau(t)|x(s),0\leq s\leq T)\sim \mathcal{N}(\mu_s(t),R_s(t)),
\end{equation}
where $t\in[0,T]$. This is the so-called smoother estimate. It differs from the filter in the sense that it exploits all the available information up to the current time instant. 
One desirable feature of the nonlinear smoother estimate in Eq. \eqref{smoother_soln} associated with the non-Gaussian system in Eqs. \eqref{model_WB} is that it can be solved via closed analytic formulae (see Appendix \ref{App_B}). We use the smoother estimate to recover the historical realizations of the two hidden variables conditioned on the entire observational period of the Ni\~no 3.4 index from 1870 to present. The results are shown in  Fig. \ref{Fig3}. Since the thermocline depth and the wind bursts are the most important precursors for predicting the state of ENSO, these recovered variables can be used to evaluate and improve prediction models and provide a guideline for improving parameterizations in coupled GCMs.


Understanding the complexity of climate dynamics presents a dilemma of completeness, between using solely statistical regression models lack the essential physics, to global climate models, which have more variables and parameterizations than observables.  Our approach presents a compromise through hybridization.  We combine a slowly-varying low-order
dynamical system with an observationally faithful statistical representation of short-time processes.  In consequence we find a realistic description of ENSO as measured by the success in (i) reproducing the observed seasonal phase locking and its uncertainty; (ii) capturing the spring predictability barrier; (iii) simulating the observed highly non-Gaussian statistics, and (iv) accurately recovering the intermittent time series of the hidden processes via a nonlinear but analytically solvable data assimilation scheme.  Furthermore, we suggest that extending the approach described here to examine climate variability on a wide range of time scales \cite[][]{GhilRMP, PinkPRL}, wherein proxy data are available, will provide important insights into mechanisms and feedbacks in the climate system.  Finally, given the generality of the stochastic dynamical systems in physics, it is hoped that our approach can be adopted not only for other problems in climate, but far more broadly in the physical sciences.



\acknowledgements

L.T.G., W.M. and J.S.W. gratefully acknowledge support from the Swedish Research Council Grant No. 638-2013-9243.  The research of N.C. is partially funded by the Office of VCRGE at UW-Madison.

\appendix

\section{Observational Data}
\label{App_A}

\begin{figure*}
\centering
\includegraphics[width=1.\linewidth]{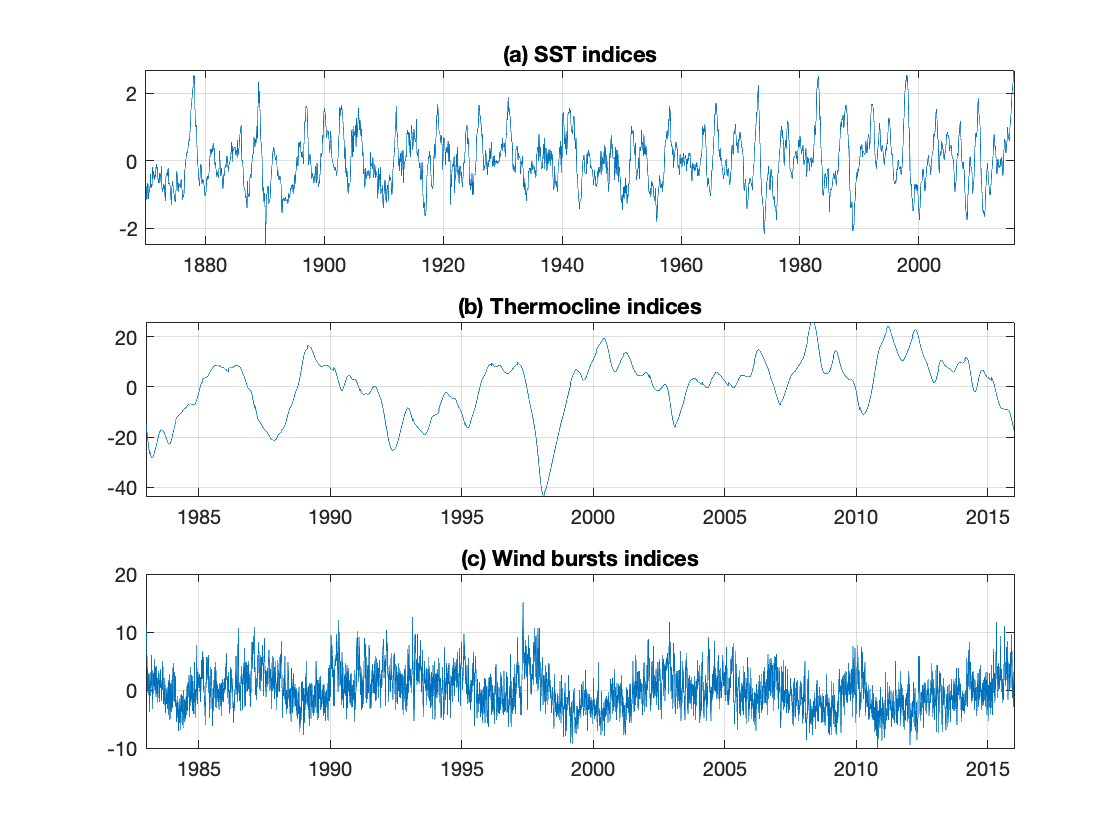}
\caption{Observational data used (a) the Ni\~no 3.4 index, (b) the thermocline depth averaged over the west Pacific and (c) the westerly wind bursts.} 
	\label{supp0}
\end{figure*}

All the observational data used in the main paper are plotted in Fig. (\ref{supp0}) and described here. 

\begin{itemize}
\item[(1)] Monthly sea surface temperature (SST) anomalies derived from the temperature analyses of the Bureau National Operations Centre (BNOC) at the Australian Bureau of Meteorology \cite{Link1}. These data have been averaged over the Ni\~no 3.4 region (5N-5S, 170W-120W) and start from the pre-industrial era spanning the entire period 1870-2016. The SST data are the only observations used in the main paper to recover the model coefficients and to reconstruct the hidden processes involved. 

\item[(2)] Daily thermocline depth data, defined as the integrated warm water volume (WWV) above the 20${}^0$C isotherm, downloaded from the NCEP/GODAS reanalysis \cite{Link2} and averaged within 5N-5S and 80W-120W in the tropical Pacific Ocean. This dataset was used in the main paper to test the performance of both the two and three dimensional models in estimating the hidden state of the thermocline depth correctly, together with their long-term (or ``equilibrium'') statistical and dynamical features, evaluated by comparing the probability distribution functions (PDFs) and the autocorrelation functions (ACFs) obtained from the models and from observations. The thermocline depth data are available only in the post-industrial era, from 1983.  Therefore, it was not possible to compare with observations the recovered hidden process over its total length, but only its realization from 1983.

\item[(3)] Daily westerly wind bursts (WWBs) at 850hPa from the NCEP/NCAR reanalysis \cite{Link3}.  By introducing the WWBs into our three dimensional model (Eqs. 2 in the main paper) we reproduced the non-Gaussian features observed in ENSO statistics that are not possible in a linear model.   The relevant observations were only available from 1983 and thus comparisons of the mean state to the statistics of the hidden 
processes recovered through the model to observations began at that date.  

\end{itemize}

\section{State Estimation and Data Assimilation}
\label{App_B}

\subsection{Online data assimilation}
Our two-dimensional model (Eqs. 1 in the main text) is a linear model with time-varying coefficients. Therefore, given the observations of the SST, the classical Kalman-Bucy filter \cite{Kalman1961} can be directly applied to obtain the state estimation of the thermocline depth.  In our three-dimensional non-Gaussian model (Eqs. 2 of the main text), the multiplicative noise prevents the use of the Kalman-Bucy filter for the state estimation. Nevertheless, conditional on the observed SST, the thermocline depth and the wind bursts remain as Gaussian processes. For such a conditional Gaussian system, the conditional distribution $p(h(t),\tau(t)|x(s),s\leq t)\sim \mathcal{N}(\boldsymbol{\mu}_f(t),\textbf{R}_f(t))$ is Gaussian. Note that despite the non-Gaussianity, the conditional Gaussian distribution can be solved via closed analytic formulae \cite{chen2018conditional}.  This facilitates an efficient data assimilation scheme to recover the states of $h(t)$ and $\tau(t)$ and their uncertainties through the filter mean $\boldsymbol{\mu}_f(t)$ and the square root of the filter covariance $\textbf{R}_f(t)$, which for our three dimensional model are given by
\begin{subequations}\label{Filter}
\begin{gather}
  \d \boldsymbol{\mu}_f = (\textbf{c}_0 + \textbf{c}_1 \boldsymbol{\mu}_f)\d t + \textbf{R}_f \textbf{C}_1^T B^{-2} (\d x - (C_0 + \textbf{C}_1 \boldsymbol{\mu}_f)\d t),\label{Filter_Mean} \\   
  \d \textbf{R}_f = \big(\textbf{c}_1 \textbf{R}_f+ \textbf{R}_f \textbf{c}_1^T + \textbf{b} \textbf{b}^T - (\textbf{R}_f \textbf{C}_1^T)B^{-2}  (\textbf{R}_f \textbf{C}_1^T) ^T\big)\d t,\label{Filter_Cov}
\end{gather}
\end{subequations}
respectively, where
\begin{equation}\begin{split}
\textbf{c}_0&=\begin{bmatrix}
-\omega x \\
0 
\end{bmatrix}, \;\;\; \textbf{c}_1 = \begin{bmatrix}
\lambda & \alpha_2 \\
0 & d_\tau 
\end{bmatrix},\\
C_0 &= a x ,\;\;\; \textbf{C}_1 = \begin{bmatrix}
\omega & \alpha_1
\end{bmatrix},\\
B&=N, \;\;\; \textbf{b}=\begin{bmatrix}
\sigma & 0 \\
0 & \rho 
\end{bmatrix}.
\label{coeffs3D}
\end{split}\end{equation}
The filter mean and covariance of the two dimensional model can be obtained from these expressions by setting $\alpha_1, \alpha_2=0$.

\subsection{Offline smoothing}
The data assimilation process exploits the observations up to the current time instant for the purpose of initialization. On the other hand, for reconstructing the unobserved variables, conditioned on the entire observational signal (such as obtaining reanalysis data), the so-called smoothing technique is preferable. For example, the smoothing of the three-dimensional non-Gaussian system aims to solve $p(h(t),\tau(t)|x(s),0\leq s\leq T)\sim \mathcal{N}(\boldsymbol{\mu}_s(t),\textbf{R}_s(t))$, where $t\in[0,T]$. The smoothing technique can be regarded as a two-step data assimilation process, with a forward pass of filtering and a backward pass of smoothing. Although there are no closed formulae for the smoother mean $\boldsymbol{\mu}_s(t)$ and covariance $\textbf{R}_s(t)$, closed analytic formulae {\em are} available for both the systems studied in the main text with appeal to the conditional Gaussian structure \cite{Chen2019}
\begin{subequations}\label{Smoother}
\begin{gather}
  \d (-\boldsymbol{\mu}_s) = \big(-\textbf{c}_0 - \textbf{c}_1 \boldsymbol{\mu}_s  + (\textbf{b} \textbf{b}^T)\textbf{R}_f^{-1}(\boldsymbol{\mu}_f - \boldsymbol{\mu}_s)\big)\d t,\label{Smoother_mean}\\
  \begin{split}
  \d (-\textbf{R}_s) =& -\big((\textbf{c}_1 + (\textbf{b} \textbf{b}^T) \textbf{R}_f^{-1})\textbf{R}_s\\&+\textbf{R}_s(\textbf{c}_1^T + (\textbf{b} \textbf{b}^T) \textbf{R}_f^{-1}) - \textbf{b} \textbf{b}^T\big)\d t,\label{Smoother_cov}
  \end{split}
\end{gather}
\end{subequations}
where $\boldsymbol{\mu}_f, \textbf{R}_f$ are the filter mean and covariance respectively, defined in Eq. (\ref{Filter}), while the coefficients $\textbf{c}_0, \textbf{c}_1,\textbf{b}$ have been specified in Eq. (\ref{coeffs3D}).

In \eqref{Smoother}, the terms of the left hand side are understood as
\begin{equation*}
\begin{split}
  \d (-\boldsymbol{\mu}_s) &= \lim_{\Delta{t}\to0} \boldsymbol{\mu}_s(t) - \boldsymbol{\mu}_s(t+\Delta{t}),\\
  \d (-\textbf{R}_s) &= \lim_{\Delta{t}\to0} \textbf{R}_s(t) - \textbf{R}_s(t+\Delta{t}).
\end{split}
\end{equation*}

\section{Parameter Estimation}
\label{App_C}

\subsection{Parameter estimation of the time-varying parameters $a(t)$ and $N(t)$}

In this section, we will use the monthly-averaged SST data to construct the periodic drift and diffusion coefficients, $a(t)$ and $N(t)$,  that appear in the $x$-equation of Eqs. (\ref{model}) in the main paper viz.,
\begin{equation}
\begin{split}
&\dot{x}(t) = a(t)x(t) + \omega(t)h(t) + N(t)\xi_x(t) ~~ \textrm{and} \\&
\dot{h}(t) = -\omega(t)x(t) + \lambda h(t) + \sigma \xi_h(t).
\label{model_app}
\end{split}\end{equation}
Here we consider a discretized version of this two dimensional model with $\Delta t$ the discrete time unit. For simplicity, we set the yearly frequency of the coefficients equal to $2\pi$. We define $N_y$ as the total number of years of the observed SST data, each of which contains  $\Delta t^{-1}$ data points. We multiply both sides of the $x$-equation by $x$, and upon taking the average we obtain
\begin{equation} 
\langle x_i x_{i+1}\rangle_i - \langle x_i^2\rangle_i = \Delta t (a_i\langle x_i^2\rangle_i - \omega_i\langle h_i\, x_i\rangle_i) \; \;\forall i \in (0,\Delta t ^{-1}],
\label{discmodel}
\end{equation}
where the average has been performed only over the quantities in the $i$-th time step inside each period $T$ with $T=1$.
Using the $h$-equation, the last term on the right hand side of Eq. \eqref{discmodel} can be written as
\begin{equation}\begin{split}
\langle h_i\, x_i\rangle_i& \simeq\frac{1}{N_y}\left(\sum_{j=1}^{N_y} \sum_{s=0}^{j \Delta t^{-1} +i\Delta t}e^{-\lambda(j\Delta t^{-1} +i\Delta t -s)}\omega_s x_s x_i \Delta t\, \right)  \\& \simeq
\frac{1}{N_y}\bar{\omega}\left(\sum_{j=1}^{N_y}\sum_{s=j\Delta t^{-1} +i\Delta t-\Lambda(\lambda)}^{j \Delta t^{-1} +i\Delta t} x_s \, x_i \Delta t \right) \simeq \\& \simeq \bar{\omega} \langle x_{i+\frac{1}{4\Delta t}}\, x_i\rangle_i \simeq 0.
\label{Deltaa}
\end{split}\end{equation}
Here we have used the fact that $x$ is periodic and oscillates around a mean of zero with an average frequency of $\pi/2$ (the natural frequency of ENSO is approximately 4 years),  which is four times smaller than that of $\omega$.  Therefore, this last coefficient has been approximated in the sum with its average over one period. 
Moreover, we have exploited the effect of the exponential inside the sum, which provides a cutoff in the lower limit. 
Thus, the average of this periodic function is a phase shift of a quarter of period and we can approximate this term as zero.

We can then write
\begin{equation}
a_i=\frac{\langle x_i x_{i+1}\rangle_i - \langle x_i^2\rangle_i}{\Delta t\langle x_i^2\rangle_i}+\Delta a_i,
\label{eq_a}
\end{equation}
where $\Delta a_i$ is the subleading correction to $a_i$ arising from the term $\langle h_i\, x_i\rangle_i$.

Taking the square of the $x$-equation in Eq. (\ref{model_app}) and then averaging, we can isolate $N_i$ viz., 
\begin{equation}\begin{split}
N_i^2 \Delta t =& \langle \left[x_{i+1}- x_i -\Delta t (a_i x_i - \omega_i h_i ) \right]^2 \rangle_i \simeq\\  \simeq &\langle \left[x_{i+1}- x_i -\Delta t a_i x_i\right]^2 \rangle_i  + \Delta t^2 \omega_i^2 \langle h_i^2 \rangle_i \\
&\forall i \in (0,\Delta t ^{-1}],
\end{split}\end{equation}
where all the terms proportional to  $\langle h_i\, x_i\rangle_i$ have been neglected.
Now we define $y_i=x_{i+1}- x_i -\Delta t a_i x_i$ and write
\begin{equation}\begin{split}
\frac{\langle y_i^2\rangle_i - \langle y_{i+1} y_{i}\rangle_i}{\Delta t}  &=N_i^2 + \Delta t (\omega_i^2 \langle h_{i+1} h_i\rangle_i -\omega_i^2 \langle h_i^2\rangle_i ) \simeq  \\ & \simeq  N_i^2 + O(\Delta t^2) \; \;\forall i \in (0,\Delta t ^{-1}].
\end{split}\end{equation}
Finally we can write
\begin{equation}
N_i=\sqrt{\frac{\langle y_i^2\rangle_i - \langle y_{i+1} y_{i}\rangle_i}{\Delta t}}+\Delta N_i.
\label{eq_N}
\end{equation}

We have used Eqs. (\ref{eq_a}), (\ref{eq_N}) to reconstruct the coefficients $a(t)$ and $N(t)$ from the data and we have plotted them in Fig. (\ref{Fig1}). From Fig. (\ref{Fig1})(a) can be noticed a good agreement between our estimation of $a(t)$ (blue) and the seasonal Bjerknes instability index as given by \cite{Stein2010} (red) as $a(t)=-1-\sin(2\pi t)$.

\begin{figure*}
	\centering
	\includegraphics[width=1.\linewidth]{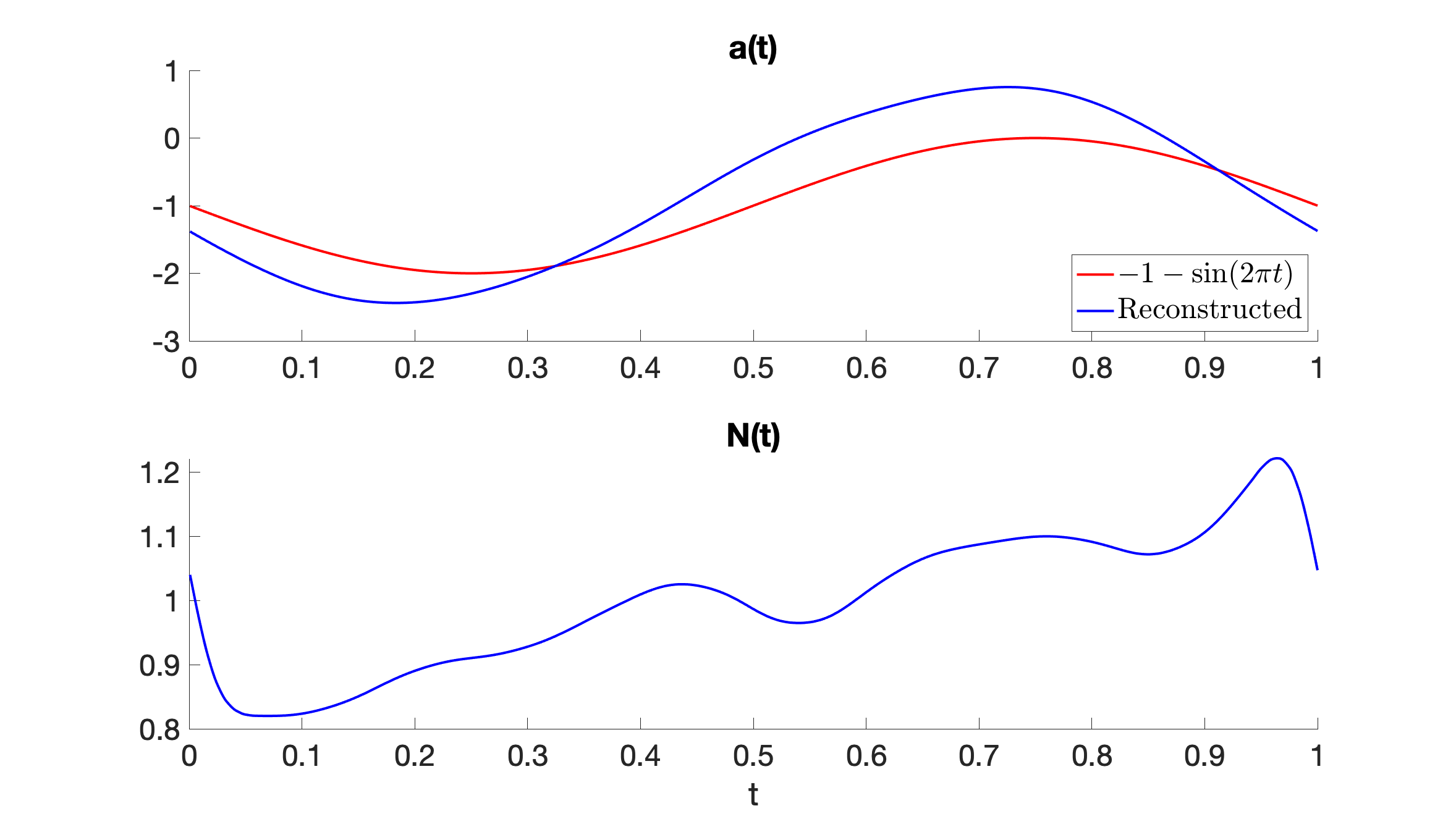}
	\caption{Shape of $a(t)$ and $N(t)$ obtained from the data (blue) and comparison with the estimation of $a(t)$ used in \cite{Stein2010} (red).}
	\label{Fig1}
\end{figure*}

\subsection{Parameter estimation of the constant coefficients using an expectation-maximization algorithm}
An expectation-maximization (EM) algorithm is used for parameter estimation of both the two- and three-dimensional systems. 
We use this algorithm because we only have the SST data over the entire observational period and hence the system is only ``partially observed''.  
Therefore, the parameters and the unobserved states of the thermocline depth and the wind bursts must be estimated simultaneously. The advantage of the EM algorithm is that it iterates and updates the parameters and the hidden states alternatively. Moreover, convergence can be guaranteed. In particular, in the E-step the smoother is applied to estimate the state while in the M-step a maximum likelihood approach is used for parameter estimation. In the E-step, we have used for convenience a normalized (i.e., with unitary variance) smoother estimate of the hidden process, which is the thermocline depth for the two dimensional model. The details of the algorithm can be found in \cite{Dempster1977}. 

Below, the EM algorithm is applied to  estimate $\omega = \omega_0 + \omega_1 \sin(2\pi t) + \omega_2 \cos(2 \pi t)$ and $\sigma$ in  the linear model.  
These coefficients have been estimated keeping the drift term $\lambda$ of the $h$-process constant in order to guarantee the stability of the algorithm. This procedure has been iterated many times for different values of $\lambda$ choosing that which minimizes both the relative and spectral relative entropy (see \cite{Hou2019} and references therein).

Figure \ref{Fig2} illustrates the iteration of the coefficients 
towards their optima; $\omega_0=1.5,\;\omega_1=0.6,\;\omega_2=-0.5$ and $ \sigma=0.9$, with the optimal value of $\lambda=-0.8$. Note that because of the simplicity of the model structure, the convergence here is global.
 





\begin{figure*}
	\centering
	\includegraphics[width=1.\linewidth]{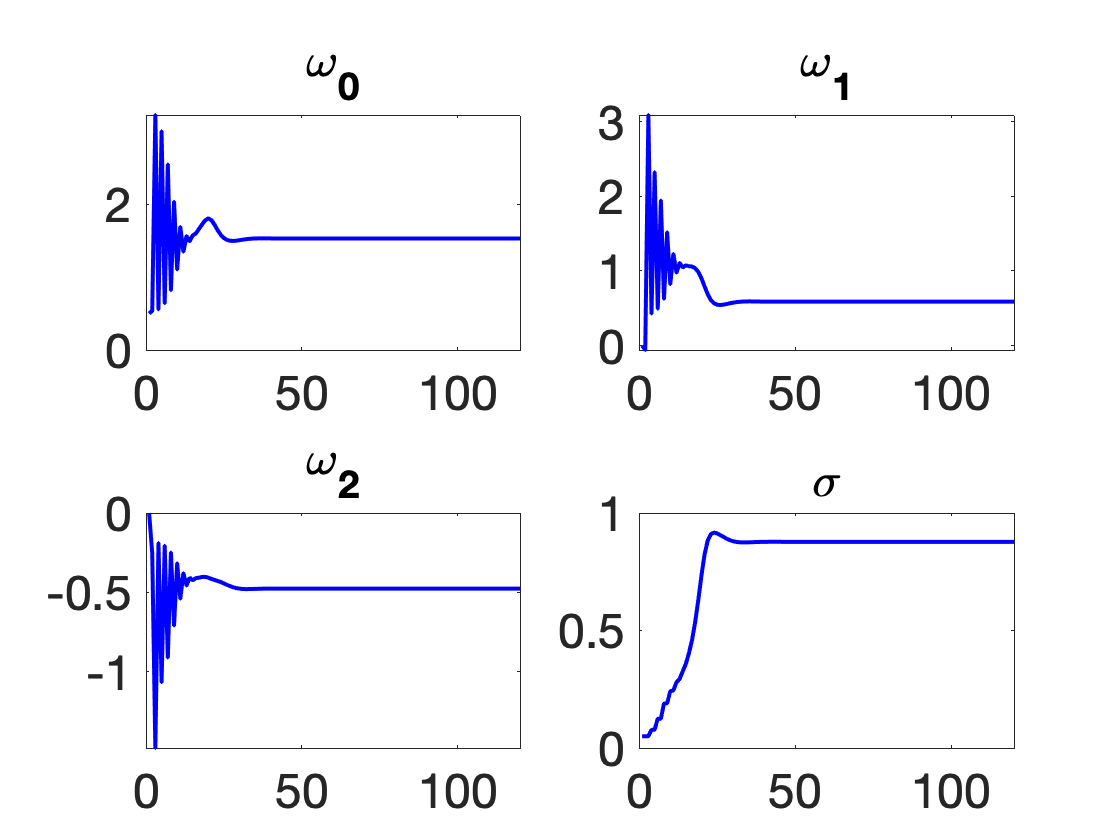}
	\caption{Coefficients of the model extracted from real data as a function of the number of iterations.  The estimated optimal values of the coefficients obtained using this method are
$\omega_0=1.5,\;\omega_1=0.6,\;\omega_2=-0.5,\; \lambda=-0.8,\; \sigma=0.9.$}
	\label{Fig2}
\end{figure*}


\subsection{Parameter estimation of the wind bursts coefficients}

We introduced a three dimensional model (Eqs. \ref{model_WB} in the main paper) to take into account the effects of wind bursts as 
\begin{equation}\begin{split}
&\dot{x}(t) = a(t)x(t) + \omega(t)h(t) +\alpha_1 \tau(t)+ N(t)\xi_x(t), \\&
\dot{h}(t) = -\omega(t)x(t) + \lambda h(t) + \alpha_2 \tau(t)+\sigma \xi_h(t) ~~\textrm{and} \\&
\dot{\tau}(t)=d_\tau \tau(t)+\rho(x) \xi_\tau(t),
\label{model_WB_app}
\end{split}\end{equation}
with the following parameters
\begin{equation}
\begin{split}
&a(t) \to 1.5a(t),\; N(t) \to 0.8 N(t),\\
&\omega_0=1.5,\;\omega_1=0.6,\;\omega_2=-0.5,\; \\&\lambda=-1.5,\; \sigma=0.8,\\
&\alpha_1=1,\; \alpha_2=-0.6, \; d_\tau=-1.5,\\& \rho(x)=4.5(\tanh(x)+1)+8.\\&\,
\end{split}
\end{equation}
The parameters that were part of the two dimensional model have been slightly modified to take into account the additional process of wind bursts.

The coefficient  $\rho(x)$ in the $\tau$ equation has two additive components.   The first is a multiplicative noise effect arising from the fact that the warmer SST in the eastern tropical Pacific is usually accompanied by strong wind bursts \cite[e.g.,][]{An2020}.  The second is an additive noise effect, which will increase the variability of the quiescent phases of the SST. These combined effects reproduce the observed non-Gaussian statistics of the model variables.

We let $d_\tau=\pi/2$, so that the damping time scale of the wind bursts is $1/4$ year, but we emphasize that this is not the time scale for the wind bursts themselves.  Rather, this is the time scale for the accumulation effect of the wind bursts, tropical convection and the MJO to trigger ENSO.  This choice ensures that the decorrelation time-scale of the wind bursts in the model is the same as that of the observations.

\end{document}